# Mechanistic Model for Deformation of Polymer Nanocomposite Melts under Large Amplitude Shear

Erkan Senses and Pinar Akcora

Chemical Engineering and Materials Science, Stevens Institute of Technology

Hoboken, New Jersey, 07030 USA

Correspondence to: Pinar Akcora (E-mail: pinar.akcora@stevens.edu)

**ABSTRACT**

We report the mechanical response of a model nanocomposite system of poly(styrene) (PS)-silica to large-amplitude oscillatory shear deformations. Nonlinear behavior of PS nanocomposites is discussed with the changes in particle dispersion upon deformation to provide a complete physical picture of their mechanical properties. The elastic stresses for the particle and polymer are resolved by decomposing the total stress into its purely elastic and viscous components for composites at different strain levels within a cycle of deformation. We propose a mechanistic model which captures the deformation of particles and polymer networks at small and large strains, respectively. We show, for the first time, that chain stretching in a polymer nanocomposite obtained in large amplitude oscillatory deformation is in good agreement with the nonlinear chain deformation theory of polymeric networks.

**KEYWORDS**    polymer nanocomposite; nonlinear rheology; deformation; networks; poly(styrene)-silica

# INTRODUCTION

Reinforced composites that are subject to frequent deformation should withstand repeated mechanical loadings and it is desired that their initial structures are retained or recovered. Therefore, it becomes critical to predict the mechanical properties of polymer nanocomposites under large deformation. Structural recovery or reformation after long deformation times adequately determines the potential applications of reinforced materials. The mechanisms for the origin of reinforcement in nanocomposites have been evaluated in terms of the state of particle agglomeration (connected particle network)[1,2] and the strength of particle-polymer interactions[3,4]. The latter effect has been proposed to be due to the immobilization of chains around nanoparticles[5,6] or due to the trapped entanglements[7] as in attractive particle-polymer composites. In the case of polymer bridging between the particles, the polymer conformation has been suggested to be an additional factor to the mechanical behavior in attractive particle-polymer networks where particles and polymer share a wetting interface[8]. Resolving these contributions of particle network, polymer bridging between particles and the percolated glassy layers to the mechanical properties has been challenging. To this end, nonlinear viscoelastic properties of composites are pertained intrinsically to the polymer matrix and are shown to enhance with the addition of fillers[1,9]. Another nonlinear work on filled elastomers shows that elastic moduli do not change with the applied large strain but modifies the lifetimes of glassy bridges[10]. The non-linearity in the elastic stress was explained by the distribution of relaxation times of the percolated glassy layers. Our work focuses on a repulsive composite system where glassy layers do not exist around the fillers. We aim to explain the origins of nonlinear contributions to the elastic stress in terms of particle structures and deformation recovery. In summary, the role of polymer bridging and trapped entanglements between nanoparticles in the reinforcement mechanism is well accepted, however previous mechanical studies are limited to observing the polymer contribution to the overall stress response[5,6,11,10]. Therefore, it is essential that rheological data should be discussed along with the structural changes under shear, but more importantly the elastic term of the stress has to be addressed for the polymer and particle parts to unequivocally distinguish their contributions as the elasticity of polymer chains and stiff materials are well known.

In this work, dynamic response from large-amplitude oscillatory shear (LAOS) experiments is discussed with the changes in particle dispersion to provide a complete physical picture of the mechanical deformation in polymer nanocomposites. LAOS experiments have been applied on microstructural systems in previous works on copolymer solutions and gels. For example, sol-gel transition of PEO-PPO-PEO triblock copolymer solutions is studied in LAOS experiments as they go through microstructural changes[12]. Likewise, the LAOS response of glutan gels was shown to be in agreement with the formation of molecular structures[13]. LAOS has also been a useful method to observe different relaxations of linear and comb polymer melts[14]. Recent improvements on the rheometers (ARES-G2, TA Instruments) for non-linear rheology capabilities as well as mathematical analysis to separate the elastic and viscous component of the total stress allowed us to resolve the nonlinearities at large strains[15,16].

We deformed the PS-Si nanocomposites (repulsive system where particles and polymer dislike each other) with large periodic strains and analyzed the elastic stress-strain data to explain the contributions of polymer and particle to the total stress for a wide range of strain. A mechanical model is developed based on a nonlinear elastic model for polymeric and particle networks to describe the rich nonlinearities. As the structure evolved during the rest time of LAOS, TEM images obtained for different states of deformation and the moduli of the corresponding states allowed us to attribute the mechanical nonlinearities to the polymer composite under tension.

**EXPERIMENTAL**

**Composite preparation.** Colloidal silica nanoparticles (13±2 nm in size) of Nissan Chemicals were mixed with poly(styrene) (52.5 kg/mol, PDI: 1.03) in 1,2 Dichlorobenzene to prepare composites containing 2, 7, 15 wt% particle loadings. The solution was first bath sonicated for 30 min, stirred rigorously for 2 h and was bath sonicated for another 30 min. The evaporation of the solvent was completed at -38 cm-Hg gauge pressure in approximately 1.5 h and then the oven was set to full vacuum (-76 cm-Hg, gauge) state. The final samples with ~250 $\mu$m thickness were annealed at 130 °C for 4 days, at 150 °C for 3 days and finally at 180 °C for 2 h under vacuum.

**Structural Characterization.** Nanocomposite bulk films were microtomed at room temperature and dispersion of nanoparticles was examined in transmission electron microscope (Philips CM20 FEG TEM/STEM) operated at 200 keV.

**Rheology.** Rheology measurements were performed on a strain-controlled ARES-G2 rheometer (TA Instruments) equipped with 8-mm diameter parallel plates geometry under nitrogen. The samples annealed according to above procedure were molded with a vacuum assisted compression molder at 150 °C for 10 min. The samples were finally annealed at 130 °C for 12 h. The measurements were performed at 170 °C following sample equilibration time of 10 min. Time sweep experiments during this period confirmed the sample stability. The large amplitude oscillatory shear (LAOS) tests were performed at a fixed angular frequency of 1 rad/s. The time dependence of the modulus was determined by fixing the strain amplitude at 500% and allowing the samples rested for increasing periods of waiting times (5-2000 sec) between 20 cycles of large deformation. Data were collected at a sampling frequency of 300 Hz, giving 1885 points per cycle. After completing each test, the samples were cooled below $T_g$ in one minute. The gap was adjusted so that the normal force never exceeds +/- 1N. This prevented deformation of the sample during the cooling step.

**RESULTS AND DISCUSSION**

Particle dispersion in all composites prior to rheology measurements was characterized in TEM. Figure 1 shows that nanocomposite at 2 wt% loading present string-like structures and further increase of the concentration to 7 and 15wt% results in percolated aggregates.

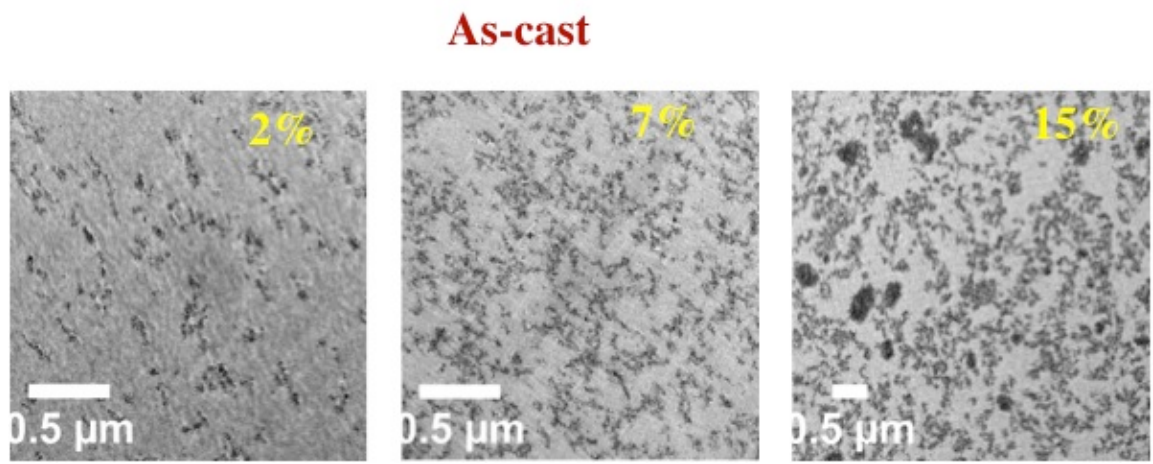

**Figure 1.** TEMs of the PS composites containing 2, 7 and 15 wt% nanoparticles

We conducted linear viscoelastic measurements on PS homopolymer to estimate the reptation time ($\tau_{rep}$) and diffusion coefficient ($D_{rep}$) of PS chains. The frequency sweep data obtained at 170 °C is shown in Figure 2a. The entanglement (plateau) modulus ($G_N$) is ~200 kPa at this temperature. The zero shear viscosity ($\eta$) is determined to be ~4000 Pa.s from the low frequency plateau of the complex viscosity ($\eta^*$) as shown in Figure 2b. The reptation time is related to $G_N$ and $\eta$ with the relationship $\eta =(\pi^2/12)\, \tau_{rep}\, G_N$. Thus, $\tau_{rep}$ is calculated to be ~0.02 sec at 170 °C. The storage moduli and loss angle in the linear viscoelastic region for homopolymer and composites at 170 °C is shown in Figure S2.

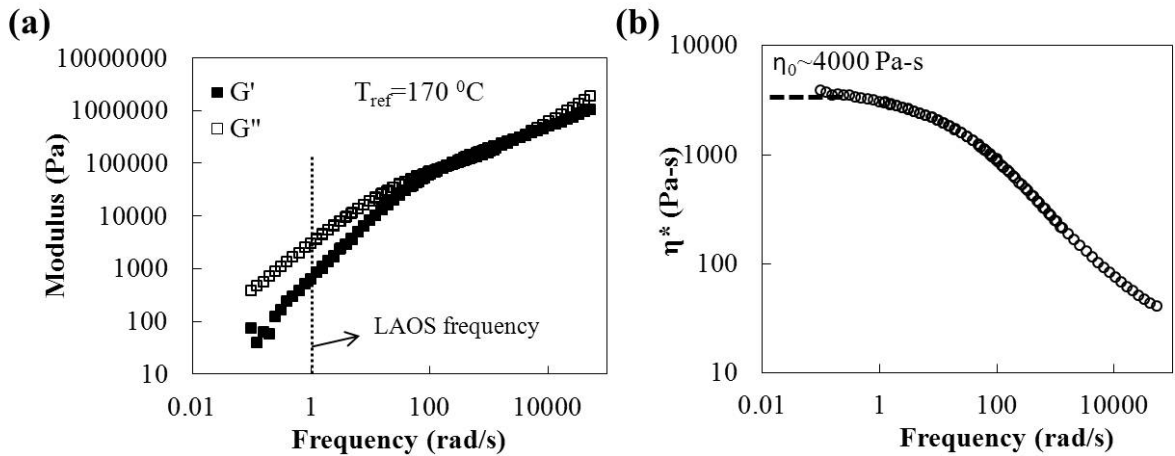

**Figure 2.** Linear viscoelastic properties of the PS homopolymer. (a) Time-temperature superposition of the storage and loss moduli for PS with reference temperature of 170 °C. Vertical line represents the frequency used in this study. (b) Complex viscosity as a function of frequency.

We strained the samples periodically in the non-linear region and observed the resulting mechanical response. When a periodic strain ($\gamma(t)=\gamma_0\sin(\omega t)$) is imposed to a sample with angular frequency, $\omega$, and strain amplitude, $\gamma_0$, the sample is strained in the forward direction from zero to the maximum strain ($\gamma_0$) in a quarter of the oscillation period, $\pi/2\omega$. In the following quarter cycle the strain is decreased to zero. This motion is then repeated in the same manner but in the opposite direction and the cycle is completed at one oscillation period. Thus, choosing the strain amplitude greater than the linear viscoelastic limit, each deformation cycle allows transition of sample from small strain (linear) to large strain (non-linear response) response.

As the samples are sheared at large-strain amplitudes, the energy stored by the stiff particle network is overcome by the external work imposed by the deformation and the chains stretching back and forth during periodic loadings may contribute to the rupture of the network. This continuous deformation is responsible for initial cycle-to-cycle strain softening (see Figure S1) which is not observed for homopolymer. Thus, the initial few cycles contain particle structural information from the previous state and we analyzed our data from the averaged stress-strain response from the initial first four cycles (see supplementary information and Figure S1). We applied 20 cycles of sinusoidal strains $\gamma(t)=\gamma_0\sin(\omega t)$, with $\gamma_0=5$ at a fixed frequency of $\omega=1$ rad/s, was necessary to reach steady stress response for any composition. The frequency was chosen to be 1 rad/s because the polymer's loss (viscous) modulus is higher than the storage (elastic) modulus at that frequency and temperature (see Figure 2a), which facilitates shearing without giving rise to instabilities such as edge fracture, wall-slip or crazing (see Supplementary Information). While the viscous component is higher than the elastic component at these conditions, the decomposition[15,18,10] of the total stress allows us to separate the elastic component as a function of resting time and concentration (see Supplementary Information for nonlinear deformation analysis). Because the deformation of stiff particle network and flexible polymer chains can contribute at different strain levels, it is important to choose strain amplitude high enough to observe both contributions in a cycle of deformation. Figure 3 shows the elastic stress-strain behavior of the homopolymer and 15 wt% loading at different strain amplitudes.

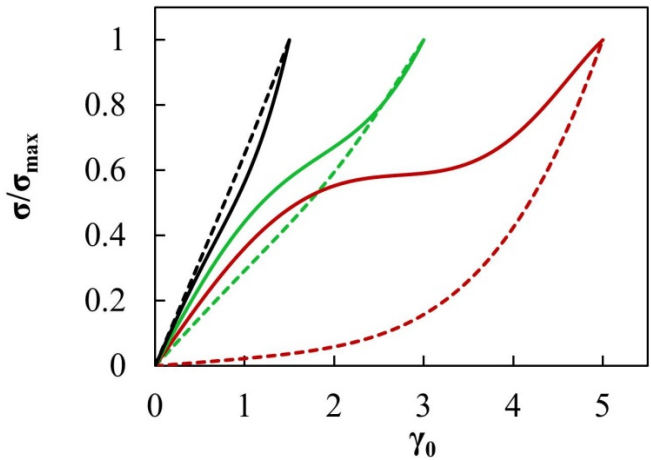

**Figure 3**. Normalized stress-strain curves of homopolymer (dashed lines) and nanocomposite with 15 wt.% compositions (solid lines) at 150% (black curves), 300% (green curves) and 500% (red curves) strain amplitudes. The measurements are done at 170 °C and 1 rad/s.

The homopolymer behaves linearly at 150% while strain hardening is observed at the largest strains when the strain amplitude is chosen to be 300% or 500%. The composite, on the other hand, shows non-linear strain softening at intermediate strains, which gets wider and exhibits clear yield-like behavior at 500% strain amplitude. The deformation pattern at this strain amplitude reveals three different regions for the nanocomposite as a function of strain; while homopolymer only gets harden at large strains without showing a yielding behavior. Thus, the strain amplitude of 5 was large enough to observe the evolution of these different deformation regimes which will be discussed in the following section.

The elastic stress-strain behavior of the homopolymer and composites are presented in Figure 4a. In order to compare composites with different particle loadings, we normalized the stress values with their corresponding maximum stress ($\sigma_{max}$). The results suggest that the stress response can be divided into three regions: At small strains (region i), the elastic stress rises linearly and the slope of the curves monotonically increases with particle concentration. This region is followed by yield-like behavior for composite samples while homopolymer's stress keeps increasing. The yield strain, which we calculated from the maximum curvature on the stress-strain curve following the linear elastic region[19] (i.e. $|d^2\sigma/d\gamma^2|$=max), was determined to be 1.60 and nearly the same for each concentration. The yield stresses (Figure 4a) of the composites show similar linear trend with the maximum stress attained in a cycle. At 15 wt% loading, highly percolated structures (see Figure 1a) create higher energy barrier for flow to occur. The elastic energy that is stored by means of particle network increases with the particle concentration, reaches to a plateau and the sample yields due to network breaking. The yielding persists over a wide range of strain since the stronger junctions require more energy to break. The specific elastic energy (energy per unit volume) to break the particle network can be estimated by E=1/2 ($\sigma_{yield} \cdot \gamma_{yield}$), which is the area under the stress-strain curves in the region (i). It is seen that the elastic energy increases with particle loading at constant yield strain. Since the aggregates at different concentrations have similar fractal-like nature, the strength of particle networks increases with the particle concentration which in turn gives rise to linear dependence of the yield stress while the yield strain becomes unchanged. The breaking of the network continues at strains between 1.6 and 3.6, where both the particles and polymer contribute to the deformation in this regime. As the strain goes beyond 3.6 the polymer gets more stiffen due to finite extensibility of the polymer chains[20]. The slope of the curve in this region does not significantly change with particle concentration, therefore deformation at this strain is attributed to the finite chain extensibility. The schematic representation of the three regions is shown in Figure 4b. The area under the normalized stress-strain curve and maximum stress determines the total elastic stress stored per volume in a quarter cycle. Thus, the particle-network both increases the strength and stored elastic energy.

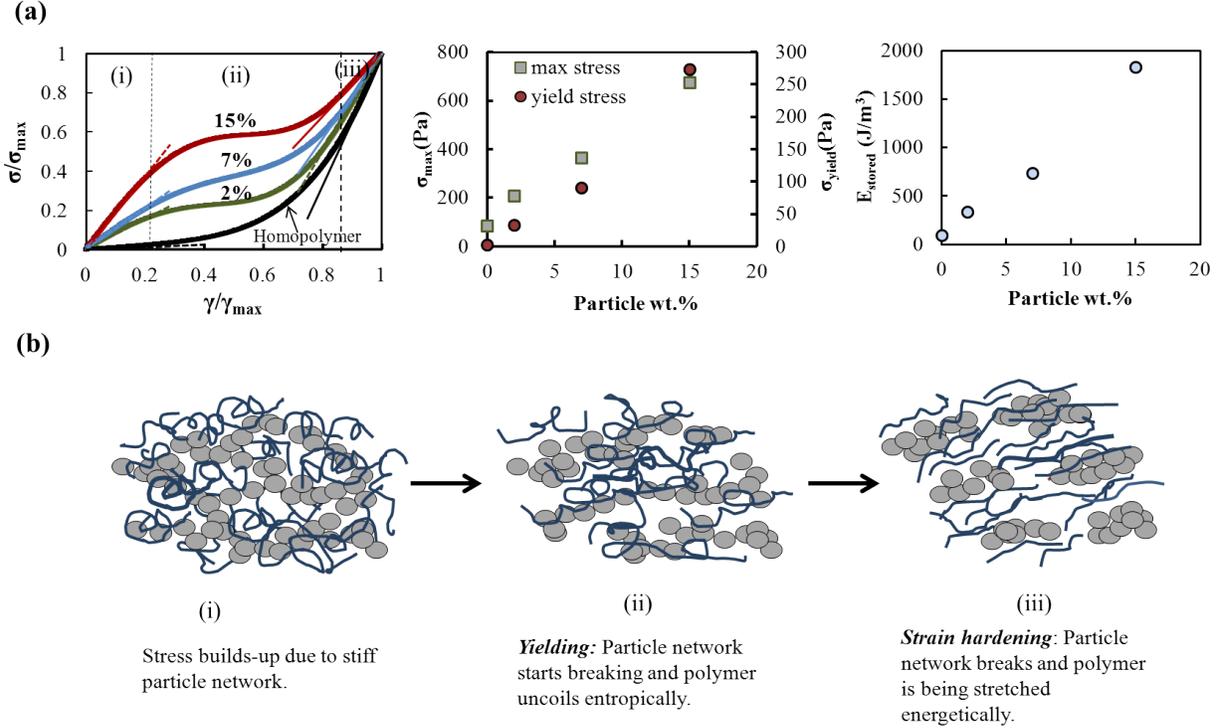

**Figure 4.** (a) Normalized stress-strain behavior of homopolymer, 2, 7 and 15 wt% nanocomposites shows the behavior of the samples at small strain; yielding and strain hardening at large strains. Maximum stress (at γ=5), yield stress and stored elastic energy are shown as a function of particle concentration. (b) Schematic representation of deformation stages of a repulsive nanocomposite having weak interactions between polymer and fillers.

Because the particles and polymer govern the deformation at small and large strain levels, respectively, we developed a non-linear deformation model that explains the elastic stress-strain behavior at a wide range of strain. Figure 5a presents the mechanical analog of the model which consists of a nonlinear elastic spring ($\sigma=G_{part}\gamma^n$) with strain softening parameter, n, that is attached to a friction element (with critical stress and strain, $\sigma_y$, $\gamma_y$) in parallel with a Gaussian coil. At strains smaller than the critical strain, only the elastic spring responses to deformation and gives rise to the initial stress build-up. Beyond the critical strain, the friction element starts moving with a constant stress $\sigma_y$, while the Gaussian coil is stretched. At large deformation (above 3.6 strain) the polymer coil is highly stretched, the chain presumably deviates from Gaussian statistics and final hardening is observed. Elasticity of a single polymer chain and polymeric networks has been well studied both theoretically and experimentally in biological systems and gels[21,20,22,23]; however, stretching of polymer chains in melt has not been directly captured from rheology due to the presence of viscous contributions to the measured stress. Since we separated the elastic component of the stress at high strains, the nonlinear shear deformation theory for polymer networks[24] apply in our case and the stress response of the polymer coil can be expressed as: $\sigma = \frac{G_{pol}\gamma}{3}\left[1 + 2\left(1 - \frac{\beta I_1(\gamma)}{3}\right)^{-2}\right]$, where $G_{pol}$ is the shear modulus of the polymer, $I_1(\gamma)=(\gamma^2+3)$ is the first invariant for the shear deformation and $\beta$ is the chain elongation ratio and defined as the ratio of the mean square of the end-to-end distance of an unperturbed chain to the square of the end-to-end distance of the fully extended chain ($\beta = \langle R_0^2 \rangle / R_{max}^2$). The overall stress functions can be expressed at different strain limits as:

$$\sigma = \begin{cases} G_{part}\gamma^n, & 0 \leq \gamma < \gamma_y \\ \frac{G_{pol}(\gamma - \gamma_y)}{3}\left[1 + 2(1 - \frac{\beta I_1(\gamma - \gamma_y)}{3})^{-2}\right] + C, & \gamma_y \leq \gamma \leq \gamma_0 \end{cases}$$

The constant C is used to ensure the continuity of the stress at the yield point and is equal to the constant yield stress generated by the friction element ($C=\sigma_y=G_{part}\gamma_y^n$). The elastic stress-strain curves for homopolymer and composites and the corresponding model fits are shown in Figures 5b and c. Note that for the homopolymer, the particle contribution does not exist. We used $\gamma_y=1.6$ for all the compositions as determined from the curvature of the stress-strain data. The strain softening parameter (n) is found to be 0.84±0.035, nearly constant for all the composites. Stiffness of the polymer chains ($G_{pol}$) is found to be increasing with particle concentration and reaches a plateau after 7% loading. In addition, extension ratio (β) of homopolymer increases from 0.06 to 0.1 with the addition of fillers, which is in agreement with the reported values of polymer networks[24]. Increase in β suggests that the chains in the composites are less stretched than homopolymer. Although the slopes are discontinuous at the critical strain, this model properly relates the macroscopic behavior to the microscopic theoretical models. Similar models exist in the literature to describe nonlinear elastic behavior of the flexible gels and biological networks[24,25,26,27,28]. For example, the spider silk presents very similar nonlinear elastic behavior where nodes acts as sacrificial elements and material presents yield-like behavior after initial linear increase[29]. The silk presents stiffening at the largest strains due to finite extensibility of the initially coil-like polymers. In that case, the atomistically derived nonlinear model was utilized where the stiffening behavior is captured with a simple exponential function. The model presented here is based on well-known nonlinear deformation theory of polymer chains and networks[24,30] and it is demonstrated here for the first time that the LAOS experiments can capture the single chain deformation.

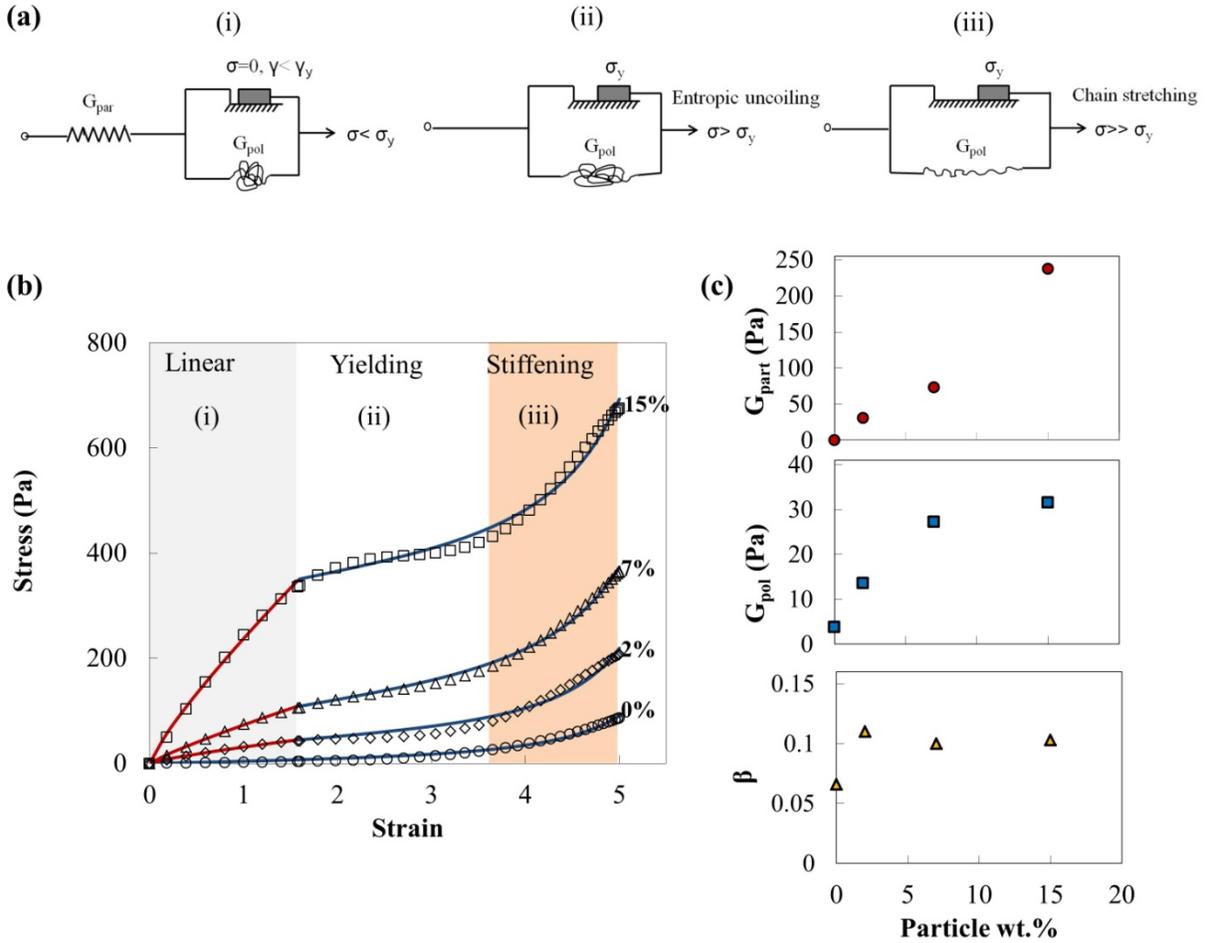

**Figure 5**: (a) Mechanical model for the elastic stress strain curves for different compositions. A power law spring ($\sigma=G_{part}\gamma^n$) with strain softening parameter, n, is combined with a friction element in parallel (with critical stress and strain, $\sigma_y$, $\gamma_y$) and a Gaussian coil. At strains smaller than the critical strain, only the elastic spring responses. Beyond the critical strain, the friction element freely moves with a constant stress $\sigma_y$ while the Gaussian coil is stretched. (b) The elastic stress-strain curves for homopolymer and composites with model fits. (c) The fitting parameters, $G_{part}$, $G_{pol}$ and $\beta$ as a function of particle loading.

In order to further elucidate the particle and polymer contributions to the elastic stress, we designed the experimental protocol as schematically shown in Figure 6a. We applied 20 cycles of deformation with the same parameters ($\gamma_0=5$, $\omega=1$ rad/s) in each deformation step. Between those steps, the samples are allowed to rest for certain times, increasing from 5 to 2000 seconds. The samples are rested for time periods larger than the relaxation time of the polymer (~0.02 sec). This allows relating the change in mechanical behavior to the particle structural evolution but not to the polymer's relaxation during rest time. The stress-strain response from the initial few cycles of each deformation step thus represents the behavior of the structure attained in the preceding rest time. Figure 6b shows the stress-strain behavior of the homopolymer and composites at initial, after 10 sec and 2000 sec resting. The yield behavior at the initial stage disappears in the intermediate stage (10 sec) and reappears after resting 2000 sec while the behavior of homopolymer presents the same nonlinear behavior at any stage. With measurements following 10 sec resting time (shown in red marks in Figure 6b and labeled as "deformed"), we probe the mechanical properties of the deformed structures since the structures cannot evolve in 10 sec. Note that the particle diffusion distance, $x\sim(Dt)^{0.5}$ is calculated as ~1 nm in 10 sec (with $D\sim8$ nm$^2$/sec, in supplementary information). It is interesting to note that the stress data collected on deformed composites after 10 sec waiting do not yield as in homopolymer. After measuring the stress response on samples deformed and aged at different times, next we look into the structures in TEM. Figure 7b shows the strong segregation of particles from the deformed polymer and waited for 10 sec waiting. This allows us to conclude that yielding, which is a result of particles restructuring, appears at small strains. After waiting 2000 sec following the deformation cycle, the stress values reach to the initial stress values of composites, which is also depicted in TEM images.

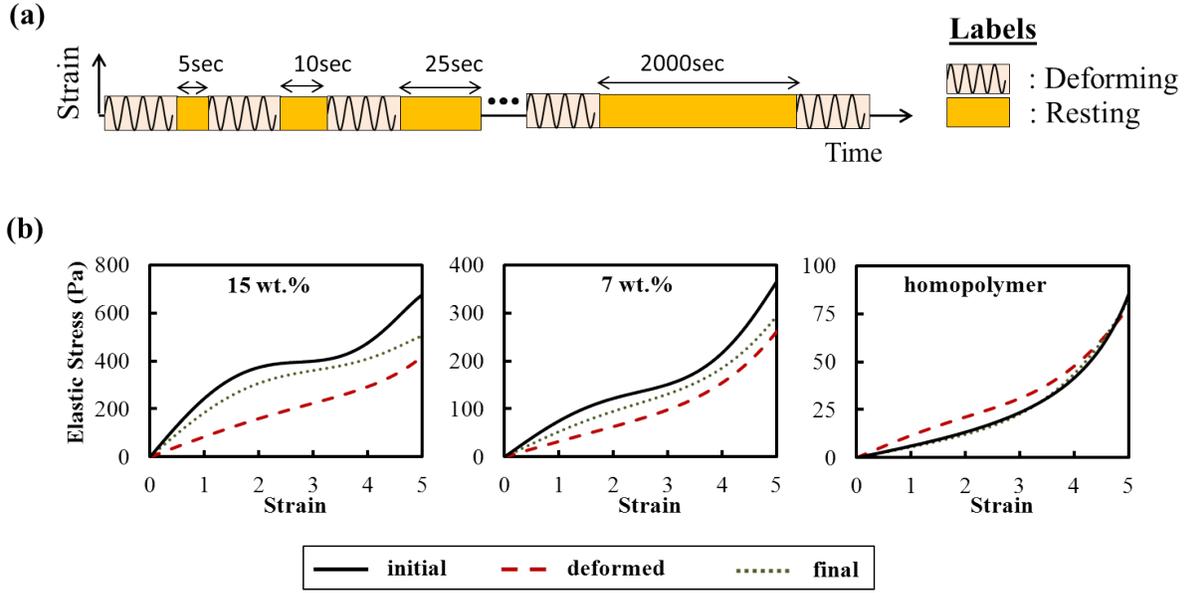

**Figure 6.** (a) Schematic representation of the experimental timeline: 20 cycles of large periodic deformations with strain amplitude of 500% and frequency of 1 rad/s in each deformation step. Samples were allowed to rest at certain time periods (5, 10, 25, 50, 100, 300, 900 and 2000 sec) between each deformation steps. (b) Stress-strain curves for homopolymer, 7 and 15 wt% compositions at different rest times: deformed (after 10 sec waiting), final (after 2000 sec waiting).

Next, we calculated the moduli at small (<1.6) and large (>3.6) strains and compared their evolution as a function of resting time in Figure 7a. It is seen that the large strain modulus ($\frac{\sigma(\gamma_0)}{\gamma_0}$) has a weak dependence on waiting time as it is mainly controlled by the matrix polymer. However, modulus at small strain ($\lim_{\gamma \to 0} \frac{d\sigma}{d\gamma}$) decreases after short waiting times particularly for 15% composite and gets close to the initial value with aging at longer times. This behavior is in-line with the particle structuring at different waiting times as discussed in previous section. At longer resting times, the large clusters break leading to an increase in small-strain modulus and yielding reemerges. Here, we comment on the size of our particles and polymer tube length. Our particles of 13nm in size are larger than the tube length of 52 kDa PS (~ 8.5 nm). In a previous work, Rubinstein[31] suggested that at times longer than the relaxation time of the polymer, particles larger than the tube diameter experience ordinary diffusion determined by the bulk viscosity of the melt. In our experiments, the reptation time was 0.02 sec, which is much shorter than the resting times. Thus, particles diffuse during long waiting times of 2000 sec. Since the rest times are longer than the reptation time of poly(styrene) at any molecular weight, similar aggregation and redispersion of particles remain valid for the same size of particles.

(a)

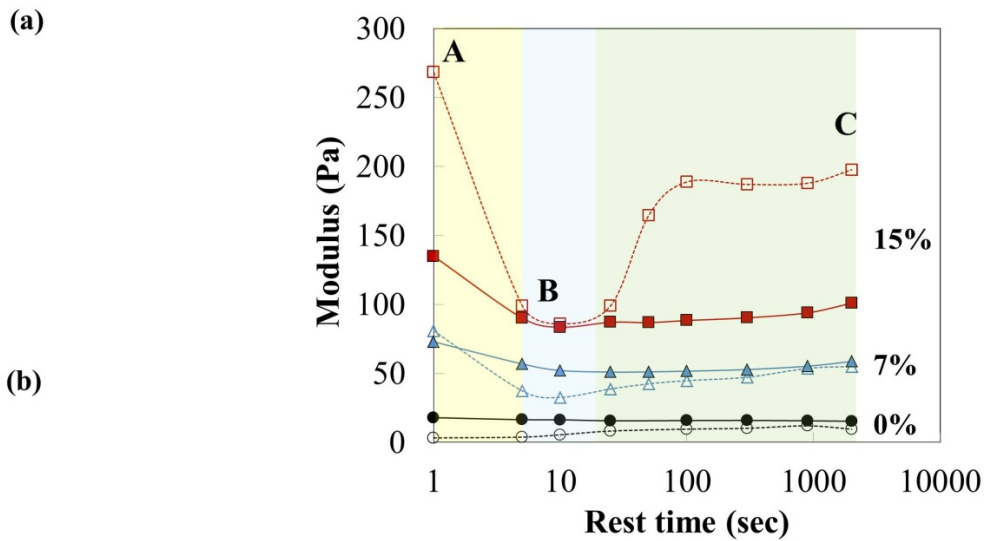

(b)

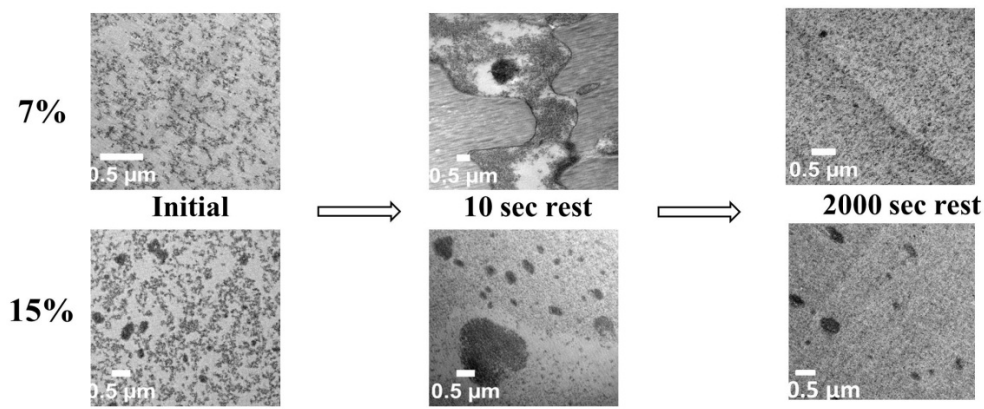

**Figure 7.** (a) Moduli calculated at small (open symbols) and large (filled symbols) strains as a function of rest time. (b) The TEM images obtained at initial (A); deformed (10 sec resting) (B); and final (2000 sec resting) (C) stages depict the network breaking and phase segregation in the intermediate stages and following re-dispersion of the particles at longer times.

**CONCLUSIONS**

In summary, we have reported the elastic stress contributions from the particle and polymer components of silica filled PS composites in large-amplitude oscillatory measurements. We show that the stress response at small strains is the result of percolated particle network whereas the large strain behavior is controlled by the finite extension of a single chain. By examining the structures of composites at different deformation states, we have shown that the reformation of initial structures is driven by the particles diffusing in liquid polymer during rest times. We have suggested a mechanistic analog for the deformation mechanism and demonstrated that the elastic stress-strain response of nanocomposites is in good agreement with the nonlinear deformation theory for the polymer that is coupled with the power law expression corresponding to the particles' deformation. Through LAOS and TEM experiment results, we conclude that when a nanocomposite is deformed at large strains, polymer chains are highly stretched around nanoparticles, and composites strain-harden as analogous to polymeric networks under deformation. The nonlinear mechanical results presented here for a repulsive composite system showcase the importance of filler-polymer interactions and filler dispersion on the elastic response of reinforced nanocomposites. Further, LAOS experiment protocols presented here can be applied to other composite systems such as polymer-grafts and attractive composites to separate the mechanical differences between bound layer existing systems and polymer-grafted nanoparticles. We conjecture that the deformation of grafted and matrix homopolymers can both contribute to the elastic stress while deformation of the bound homopolymer around particles will be rather different and possibly require more complicated modeling which will be the focus of future studies.

## ACKNOWLEDGMENTS

We thank the Stevens Institute of Technology for start-up funds and acknowledge NSF-CAREER award grant #1048865-DMR. We also thank Dr. Randy Ewoldt for sharing the MITLaos software.